\documentclass[pre,showkeys,amsmath,nofootinbib,floatfix]{revtex4}
\usepackage{graphicx} 
\usepackage{graphics} 
\usepackage{psfrag} 
\usepackage{epsf} 
\begin{document}

\title{Scaling functions for nonequilibrium fluctuations:\\
                        A picture gallery}

\author{Zolt\'an R\'acz}

\email{racz@poe.elte.hu}

\affiliation{Institute for Theoretical Physics, E\"otv\"os University, \\
P\'azm\'any s\'et\'any 1/a, 1117 Budapest, Hungary}



\begin{abstract}
The emergence of non-gaussian distributions for macroscopic
quantities in nonequilibrium steady states is discussed with
emphasis on the effective
criticality and on the ensuing universality of distribution
functions. The following problems are treated in more detail:
nonequilibrium interface fluctuations
(the problem of upper critical dimension of the Kardar-Parisi-Zhang
equation), roughness of signals displaying Gaussian 1/f power
spectra (the relationship to extreme-value statistics),
effects of boundary conditions (randomness of the
digits of $\pi$).
\end{abstract}


\maketitle

\section{INTRODUCTION}
\label{sect:intro}

Spatially averaged (global) quantities in homogeneous
equilibrium systems display Gaussian fluctuations since
the correlation length is, in general, finite and the central limit theorem applies.
The situation is more complicated at critical
points where the long-range correlations yield diverging fluctuations which, in turn,
generate nontrivial probability distribution
functions (PDF-s). A remarkable simplification occurs, however, even in
critical systems where the large-scale fluctuations lead to
{universality \cite{{Chaikin-bk},{SK-Ma-bk}}. For  the PDF-s of
macroscopic quantities, this means that the shape of the PDF-s depends only
on a few general characteristics of the system (dimension, symmetries,
range of interactions) \cite{Bruce81,Binder81}.
The emergence of universality
from large-scale fluctuations appears to be so robust that one expects
that similar mechanisms works in far from equilibrium steady states as well.
Since no general theory exists for nonequilibrium systems, studying the
differences between equilibrium and nonequilibrium phase transitions
can indeed be instrumental in understanding some distinguishing
but still robust properties of nonequilibrium
{systems~\cite{SchmZia-bk,Marro-Dick-bk,Racz-LH}}.
In particular, it may help making inroads in the largely unknown
territory of nonequilibrium PDF-s.

At first sight, universality ideas should be of restricted use
since they apply only to critical points. One should remember, however,
that nonequilibrium systems displaying power law behavior in their
various characteristics (correlation in space
or time, fluctuation power spectra, size-distributions, etc.)
are abundant in nature. Examples range from
interface fluctuations \cite{Krugreview} and
dissipation in turbulent systems \cite{turbulence}
to voltage fluctuations in {resistors \cite{Weissman}},
and to the number of earthquakes vs. their {magnitude
\cite{earthquakes}}. The underlying reason
for the ubiquity of power-laws is not understood (the
widely used expression {\it self-organized criticality} \cite{SOC-Bak}
is a testimony for this fact) but the observed effective criticality
suggests that a classification
of nonequilibrium PDF-s can be developed
using the logics of critical phenomena. Namely,
strong fluctuations and power-law correlations imply universal
scaling functions for the PDF-s and, consequently, the nonequilibrium
PDF-s in a large number
of phenomena can be determined by studying the
nonequilibrium universality classes.

Compared with equilibrium systems,
complications are expected to arise from the fact that the properties
of nonequilibrium steady states are determined not only by the
interactions but by the dynamics as well. Thus,
dynamical symmetries (conservation laws, the effects of breaking of time-reversal
symmetry, etc.) should also play an important role in the classification. Furthermore,
in spite of being universal, the scaling functions do depend on the boundary
{conditions \cite{PrivFSS}}. Although this is an extra complication, it indicates that
the scaling functions may be suitable for describing an important feature of
nonequilibrium states, namely, that the bulk behavior depends on
the boundary conditions (note that fluxes are often generated by an appropriate
preparation of the boundaries).

General (field-theoretic) studies of nonequilibrium universality
classes~\cite{SchmZia-bk,Grinstein-1993,Uwe-2002} do not
address the question of distribution functions due to technical difficulties.
For practical purposes, on the other hand, it is important to have the
scaling functions associated with the PDF-s since, as we shall see below,
they provide a possibility for "fit-free" comparisons with experiments.
Once a gallery of such scaling
functions has been built, it can be used to
identify symmetries and underlying dynamical mechanisms in experimental systems;
to discover analogies between seemingly different systems due to both
belonging to the same universality class, and
the applications are restricted only by imagination.

The project of building the picture gallery
of scaling functions associated with nonequilibrium PDF-s has
been going on (perhaps unknowingly) for several years.
For simple systems, the scaling functions have been found
{analytically~\cite{{FORWZ},{PRZ94},{AR96},HoldsXY1,1f,1falpha,Eisler2003}}.
In most of the cases, however,
the PDF of the "microscopic" configurations is unknown and thus one has
to resort to other means of calculations. One of them  is a phenomenological
approach \cite{RP94,Rosso2003,Doussal2003} which consists of introducing an effective
Gaussian action with singular dispersion and fixing the dispersion to yield the observed
scale-invariant fluctuations. The other is the brute force simulations
of models which are believed to be in the same universality class as
the system at {hand \cite{KPZ-width,HoldsPRL2000}}. The resulting picture gallery
is far from complete but contains a few interesting pieces which will be presented
below. First, I will show how the scaling functions are defined and
calculated using simple examples from surface growth problems
(Sec.\ref{sect:Surfacegrowth}). The details of the derivation will be demonstrated
on the example of a one dimensional ($d=1$) surface dynamics  that is
equivalent to the problem of Gaussian $1/f$ noise (Sec.\ref{sect:1f}).
As we shall see, this calculation establishes a connection between the
the $1/f$ noise and one of the limiting distributions of extreme statistics.
Applications of the scaling functions will be discussed in Sec.\ref{sect:KPZ}
with details presented in connection with the problem of the
upper critical dimension of the Kardar-Parisi-Zhang (KPZ) equation.
Finally, as a demonstration of the importance of the boundary conditions, we
shall discuss the problem of the randomness of the digits
of $\pi$ (Sec.\ref{sect:pi}).

\section{Surface growth and scaling functions}
\label{sect:Surfacegrowth}

Among the nonequilibrium systems displaying "effective" criticality,
growing surfaces provide a conceptually simple and
versatile laboratory from both experimental and theoretical point of
{view\cite{Krugreview,Barabasi}}.
The criticality here means that growing surfaces are usually rough i.e.
the mean-square fluctuations of the interface diverge with system size.
   \begin{figure}[ht]
   \begin{center}
   \vspace{0.2truecm}
    \begin{tabular}{c}
   \includegraphics[height=5.5cm]{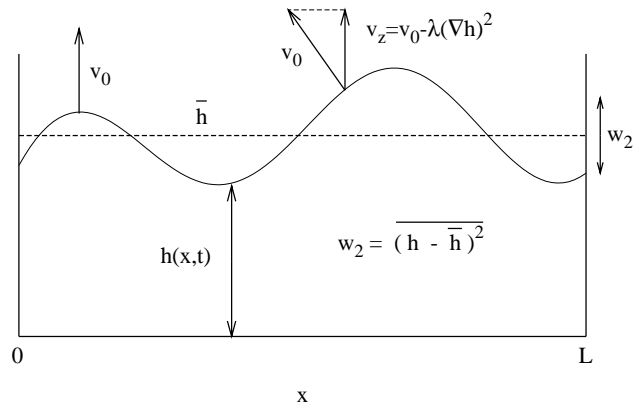}
   \vspace{-0.3truecm}
   \end{tabular}
   \end{center}
   \caption[]
   { \label{fig:surf-growth}
   Surface formed by deposition and evaporation. The vertical velocity of the surface
   $v_z=\partial_t h(x,t)$ is assumed to depend on the local properties,
   $v_z(\partial_xh,\partial_{x}^2h,...)$ and, at a macroscopic level of description,
   only a low-order gradient expansion of $v_z$ is kept.
   The growth models differ from each other
   by the physical processes which are considered to be relevant and
   thus by keeping only the appropriate terms in the gradient expansion.
   }
   \end{figure}
More precisely, let us denote
the height of the surface above a $d$ dimensional
substrate of characteristic linear dimension $L$ by $h(\vec r,t)$ (see Fig.\ref{fig:surf-growth}).
Then the mean-square fluctuations are defined as
\begin{equation}
w_2=\frac{1}{A_L}\sum_{\vec r} \left[\,h({\vec r},t)-\overline{h}
\,\right]^2\,
\label{widthsq}
\end{equation}
where the spatially averaged height is given by
${\overline h}=\sum_{\vec r}h({\vec r},t)/A_L$
with $A_L$ being the area of the substrate. For rough surfaces one finds
that the steady state average of the fluctuations $\langle w_2\rangle_L$ diverges as
$\langle w_2\rangle_L\sim L^{2\chi}$,
and the critical exponent $\chi$ is, in principle, a characteristic of the universality
class the growth process belongs to.

The divergence of $\langle w_2\rangle_L$ suggests that
the width (or roughness) distribution $P_L(w_2)dw_2$,
defined as the probability that $w_2$ is in the interval $[w_2,w_2+dw_2]$,
is a natural choice when searching for nontrivial distributions.
Indeed, if the picture about criticality is correct then $\langle w_2\rangle_L$ gives the
only relevant scale in the problem and thus it follows from dimensional
analysis that $P_L(w_2)$ can be expressed in the following form
\begin{equation}
 P_L(w_2)\approx \frac{1}{\langle w_2\rangle_{{}_L}} \Phi \left(
 \frac{w_2}{\langle w_2\rangle_{{}_L}} \right)
 \label{Phi}
\end{equation}
where $\Phi(x)$ is a universal scaling function, the object of our main interest in this
talk.

From a utilitarian point of view, three questions should be immediately answered. First,
can this quantity be measured in experiments; second, can we calculate it theoretically;
and third, are the $\Phi(x)$-s of different universality classes sufficiently different
to be distinguishable. The answer to all three questions is a yes. Present day experiments
can measure surfaces at high resolution \cite{exp94-1,exp94-2} and thus
the $P(w_2)$ distribution can be {built \cite{RP94}}. As to the theoretical
calculation, if a model can be simulated then $P(w_2)$ can, of course, be measured.
In simple cases, however, one may know the nonequilibrium steady-state distribution
${\cal P}[h(\vec r)]\sim \exp\{-S[h]\}$
and $\Phi$ can be calculated exactly (an example will be presented in Sec.\ref{sect:1f}).
Here we sketch only the first steps of the calculation.

Formally, $P(w_2)$ is obtained from ${\cal P}[h(\vec r)]$ as a path integral
\begin{equation}
P(w_2)  =  \int {\cal D}h(\vec r) \,
\delta\left( w_2 - [\, \overline{h^{\, 2}} \,-\, \overline{h}^{\, 2} \, ]
\right)
\exp\{-S[h]\}\ .
\label{Pw_2}
\end{equation}
where the overbar $\overline {h^n}$ denotes spatial averaging.
In practice, it is easier to calculate the generating function
\begin{equation}
G(s)=\int_0^\infty e^{-s w_2} P(w_2)dw_2={\cal N} \int {\cal D}h(\vec r) \,
\exp\{{-S[h]-s[\, \overline{h^{\, 2}} \,-\, \overline{h}^{\, 2} \, ]}\}\ ,
\label{Glambda}
\end{equation}
where $\cal N$ is a normalization constant. The above expression is
instrumental in finding models where
$P(w_2)$ can be obtained analytically. The path integral in Eq.(\ref{Glambda})
is the partition function of a model with an effective action
$S_{eff}[h]=S[h]+s[\, \overline{h^{\, 2}} \,-\, \overline{h}^{\, 2} \, ]$.
The terms added to $S[h]$ are quadratic functionals of $h$ and so
one expects that
the generating function and thus $P(w_2)$ can be evaluated exactly if the
original model defined by $S[h]$ is solvable.
   \begin{figure}[ht]
   \begin{center}
   \begin{tabular}{c}
   \includegraphics[height=6cm]{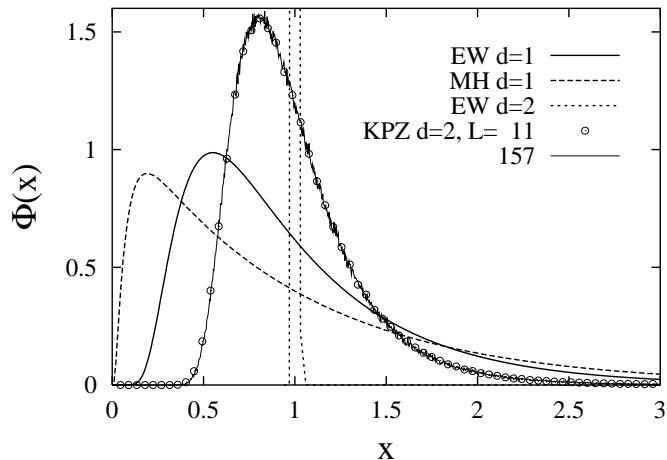}
   \vspace{-0.5truecm}
   \end{tabular}
   \end{center}
   \caption[example]
   { \label{fig:EW-MH-KPZ}
   Fluctuation distributions for the $d=1$ dimensional Edwards-Wilkinson (EW)
   and Mullins-Herring (MH) models, and for the $d=2$ EW and Kardar-Parisi-Zhang (KPZ)
   models. The scaled probability of the roughness $\Phi=\langle w_2\rangle P(w_2)$ is plotted
   against the scaled roughness $x=w_2/\langle w_2\rangle $. The $d=1$ scaling functions are the
   limiting functions when the size of the system $L$ goes to infinity. For the $d=2$
   EW model, a large but finite-$L$ scaling function is plotted since one obtains a
   delta function in the $L\to\infty$ limit. The KPZ distributions were built by simulating
   $L\times L$ systems.}
   \end{figure}
There are several growth models which are exactly solvable since
$S[h]$ is a quadratic functional of $h$. A notable example is
the Edwards-Wilkinson (EW) model \cite{Krugreview} describing growth governed by
surface tension and having $S[h]\sim \overline{ (\nabla h)^2}$.
Another example is the Mullins-Herring (MH) equation \cite{Krugreview}
modeling curvature driven growth and having a steady state characterized by
$S[h]\sim \overline{ (\Delta h)^2}$. These models belong to distinct universality
classes since $\langle w_2\rangle_L\sim L^{2\chi}$ with $\chi_{EW,d=1} =1/2$
while $\chi_{MH,d=1} =1$ in $d=1$ dimension. Accordingly, their scaling
functions should be different.
The $d=1$ scaling functions for these
models \cite{FORWZ,PRZ94,RP94,HoldsXY1} are displayed on Fig.\ref{fig:EW-MH-KPZ}
and one can see that they are easily distinguishable.

Fig.\ref{fig:EW-MH-KPZ} also shows
the $\Phi$-s for the $d=2$ EW and KPZ models \cite{RP94,HoldsXY1,KPZ-width}
(the latter is a nonlinear model discussed in Sec.\ref{sect:KPZ}) which are
again in different universality classes since $\chi_{EW,d=2}=0$ (logarithmic divergence) while
$\chi_{KPZ,d=2}\approx 0.39$. As can be seen from Fig.\ref{fig:EW-MH-KPZ} the scaling
functions strongly differ from each other and, furthermore, they are also distinct from
the $\Phi$-s of the $d=1$ models. Thus Fig.\ref{fig:EW-MH-KPZ} gives an "answer by
example" to the third utilitarian question posed above.

We conclude this section with two notes on Fig.\ref{fig:EW-MH-KPZ}. First, it should be
remarked that the $d=2$ EW scaling function approaches a delta function
on the scale of $w_2/\langle w_2\rangle$, meaning that the fluctuations of $w_2$ do
not diverge in the $L\to \infty$ limit. The delta function, however, hides an interesting
structure which can be revealed by an appropriate choice of the
scaling {variables \cite{HoldsXY1}}. Second, the collapse of the $L=11$ and $L=157$
KPZ results demonstrates an important point about the explicite $L$-dependence
of the $\Phi$-s. Indeed, the scaling functions have
finite size corrections, i.e. they have
explicite $L$-dependence in addition to the $L$-dependence through
the argument $x=w_2/\langle w_2\rangle$. Not much is known about the
finite-size correction of $\Phi$-s but the experience with a large number of models
indicate that the explicite $L$-dependence is negligible when the
number of surface sites becomes smaller than the number of sites
in the bulk. This is what we see in Fig.\ref{fig:EW-MH-KPZ} on the example of the
$d=2$ KPZ model.

\section{Applications}
\label{sect:appl}

\subsection{Discovering structures and similarities in experiments and in simulations}
\label{sect:exp}
A straightforward way of applying the picture gallery built for surface models
is to take the results of a surface growth experiment,
build the scaling function of the width distribution,
and compare it with the pictures in the gallery. If one finds agreement with
one of the pictures then one can reason about the physical processes which are
relevant in the given growth process. These types of arguments can be found
for example in Ref.\cite{RP94}.

A more sophisticated application was
the establishment of a connection between the dissipation
fluctuations in a turbulence experiment
and the magnetization fluctuations in the $d=2$ XY model at low {temperatures \cite{BHP98}}.
Since the low-temperature fluctuations in the $d=2$ XY model are equivalent to
the surface fluctuations $d=2$ Edwards-Wilkinson model, the discovery of the above connection
prompted a search for an interface interpretation
of the dissipative structures in the turbulent {system \cite{Goldenfeld}}.

In more theoretical applications, the scaling functions developed for surfaces have
been used to find the universality class of massively parallel algorithms and thus
to establish their {scalability \cite{Korniss}}. Furthermore, they were also instrumental
in establishing the universality class of fronts propagating into unstable
{phases \cite{Tripathy}}. Below, we describe in detail two more theoretical
applications. The first is notable for its logics of approaching a controversial
problem while the second is remarkable for the puzzle in the end result.

\subsection{Upper critical dimension of the Kardar-Parisi-Zhang model}
\label{sect:KPZ}
A nontrivial feature of the approach advocated in this talk is that the building
of the PDF-s does not involve any fitting or other
procedures involving subjective judgment. In order to appreciate the advantages of this
feature, we shall discuss below the problem of upper critical dimension of the
KPZ equation.

The KPZ equation \cite{KPZ-orig} is the simplest nonlinear generalization of the
EW model. In addition to the surface-tension effects, it also takes into
account that the surface grows along its normal provided
the attachment dynamics is isotropic.
Then, as one can see from Fig.\ref{fig:surf-growth},
the vertical velocity of the surface
has a correction term proportional to $(\nabla h)^2$,
and the equation, in lowest order
in the nonlinearities, can be written as
\begin{equation}
\partial_t h = \nu {\vec \nabla}^2 h +
\lambda ({\vec \nabla}h)^2 +\eta \, .
\label{KPZeq}
\end{equation}
Here $\nu$ and $\lambda$ are
parameters, while $\eta({\vec r},t)$ is a Gaussian white noise.
The above equation has been investigated intensively since
it gives account of a number of
interesting phenomena (Burgers turbulence, directed
polymers in random media, etc.).
Nevertheless, a number of issues remained unsolved. In particular,
there is no agreement on upper critical dimension ($d_u$)
above which a mean-field theory would be valid.
Mode-coupling and other
phenomenological theories (see Ref.\cite{modecoup} and references therein)
suggest that $d_u=4$ while all the numerical work (see e.g. Ref.\cite{KPZ-num})
fail to find a finite $d_u$. There are, of course, problems with the approximations
in the phenomenological theories as well as with the multiparameter fits
of the simulation results, and the debate is rather controversial.
We show now, following the lines of our recent work \cite{KPZ-width}, how the scaling
functions of the roughness can shed some light on this controversy.
   \begin{figure}[ht]
   \begin{center}
   \vspace{1truecm}
   \begin{tabular}{c}
   \includegraphics[height=6cm]{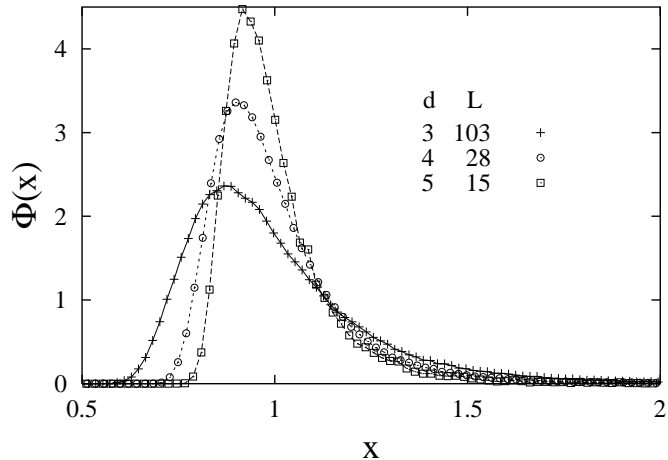}
   \end{tabular}
   \vspace{-0.5truecm}
   \end{center}
   \caption[example]
   { \label{fig:kpz3-5}
   Scaling functions for the roughness distributions in the $d=3-5$ dimensional KPZ models.
   The scaled probability of the roughness $\Phi=\langle w_2\rangle P(w_2)$ is plotted
   against $x=w_2/\langle w_2\rangle $. The characteristic linear size of the simulated
   systems is given by $L$.}
   \end{figure}

The logics of the solution originates in the theory of critical phenomena. It is
known that the universal scaling functions depend on $d$ for $d\le d_u$ while
their dependence
is lost in the mean-field region ($d>d_u$). Thus finding smoothly
varying $\Phi$-s
in dimensions $d-1, d , d+1$ should lead to the conclusion that $d\not= d_u$.
To use this argument,
we compare on Fig.\ref{fig:kpz3-5} the $d=3,4,5$ KPZ scaling functions
using simulation data for
restricted solid-on-solid (RSOS) growth {models \cite{KPZ-num,KPZ-width}}.
The message from Fig.\ref{fig:kpz3-5} is clear: It is highly unlikely that
$d_u=4$ would be the upper critical dimension of the KPZ equation.
Note that we arrived at this result without using any approximation or
fitting procedure. The only way out of the conclusion $d\not= d_u$ would be
if there would be some finite-size corrections to the scaling functions
which persist at large sizes. One cannot see such
corrections for $d=2$ in Fig.\ref{fig:EW-MH-KPZ} and
the finite-size analysis of the $\Phi$-s in $3\le d\le 5$ leads
to a similar {conclusion \cite{KPZ-width}}.

\subsection{$1/f$ noise and the Fisher-Tippett-Gumbel distribution}
\label{sect:1f}

Studies of the scaling functions often yield connections which
are intriguing but hard to understand. The aim of this section
is to derive an example of such puzzling result. Namely, we shall show
that the PDF of the mean-square fluctuations in
$1/f$ noise is related to the Fisher-Tippett-Gumbel distribution which
is one of the limiting distributions of extreme value statistics. The derivation
in itself is worth going through since it demonstrates the steps of
calculating $P(w_2)$ in a case where the usual choice of the scaling
variable $w_2/\langle w_2 \rangle$ would yield a delta-function
PDF in the large-size limit. The derivation is an expanded version
of what can be found in a recent letter of {ours \cite{1f}}.

$1/f$ noise is usually associated with a time signal so we should start
by imagining the interface in Fig.\ref{fig:surf-growth} as a time signal
i.e. we should make the replacements  $x\to t$ and $L\to T$ with $T$
being the period of the signal. The mean square fluctuations
$ w_2(h)=\overline{{[\, h(t)-{\overline{h}}\, ]}^2}$
are given by averaging over period $[0,T]$. Then the expressions
for $P(w_2)$ and for the generating function $G(s)$
[Eqs.(\ref{Pw_2}) and (\ref{Glambda})] are unchanged and one is left with the
problem of finding a suitable $S[h]$ representing the $1/f$ noise.
In Ref.\cite{1f}, we proposed
that the path probability of a Gaussian, periodic, random
phase, and perfectly $1/f$ noise with the dispersion being linear
for all frequencies, can be described by the following action
\begin{equation}
S=\sigma \sum_{n=-N}^{N} \vert n \vert \vert c_n\vert^2
=2\sigma \sum_{n=1}^{N} n \vert c_n\vert^2\, ,
\label{effham}
\end{equation}
where $\sigma$ is an effective surface tension in the language of
surfaces and the $c_n$-s are the
Fourier coefficients of the periodic [\ $h(t+T)=h(t)$\ ] signal
\begin{equation}
h(t) = \sum_{n=-N}^{N} c_{n} e^{2 \pi i n t/T}\,,\hspace{5mm}
c_{-n} = c_{n}^{\star} \ .
\end{equation}
Several notes are in order here. First, there is a cutoff in above Fourier
series. Its meaning is that the time signal is resolved at time increments
of $\tau=T/N$. Using $N$ finite makes the steps of the calculation
simple and the $N\to\infty$ (equivalent to the $T\to \infty$
limit) can be taken in the final results.
Second, the power spectrum with the above action is indeed $1/f$
\begin{equation}
\langle |c_n|^2\rangle \sim 1/|n| \, ,
\label{powspect}
\end{equation}
and, third, there is a simple meaning to $w_2$ in the noise terminology
since
\begin{equation}
w_2=2 \sum_{n=1}^{N} \vert c_n\vert^2 \, ,
\label{intpow}
\end{equation}
i.e. $w_2$ is the integrated power spectrum.

Turning now to the evaluation of $P(w_2)$, let us note
that the functional integral (\ref{Glambda})
can be written in terms of the Fourier amplitudes as
\begin{equation}
G(s) =  {\bar{\cal N}}\prod_{n=1}^{N}
\int\limits_{-\infty}^\infty dc_n \int\limits_{-\infty}^\infty dc_n^* \,
\exp\left[- \sum_{m=1}^{N}
2(\sigma  m + s) \vert c_m\vert^2\right ] \, ,
\label{Gs2}
\end{equation}
where $\bar {\cal N}$ is a constant to be determined
from the normalization ($G(0)=1$) condition. Carrying out the integrals, one finds
the generating function in terms of a product
\begin{equation}
G(s) = \prod_{n=1}^{N} \frac{\sigma n}{\sigma n + s}\ .
\label{Gsprod2}
\end{equation}
The structure of the above expression is common to various Gaussian
growth models where $S[h]$ is a quadratic functional of $h$. For example,
to obtain the $G(s)$ for the EW model, one should just make the substitution
$n\to n^2$ and $\sigma \to \sigma/N$. Since $G(s)$ has poles on the negative
real axis, there is a straighforward method for calculating $P(w_2)$. Namely,
the inverse Laplace transform of $G(s)$ is an integral along the imaginary axis,
\begin{equation}
P(w_2)= \int\limits_{-i\infty}^{i\infty} \frac{ds}{2 \pi i}
\,e^{w_2 s} \prod_{n=1}^{N} \frac{\sigma n}{\sigma n + s}
\label{Pwsc1} \, ,
\end{equation}
and thus $P(w_2)$ is obtained as a sum of contributions from the poles.
The method works \cite{FORWZ,PRZ94,RP94} but in our case
it yields a rather complicated expression which can be shown to converge to
a delta function if the usual scaling variable $x=w_2/\langle w_2\rangle$ is used.
Actually, this result can be understood by just calculating the first two
cumulants of $w_2$. The first cumulant diverges for $N\to \infty$ as
\begin{equation}
\langle w_2 \rangle = \left. - \frac{dG}{ds} \right|_{s=0}
= \frac{1}{\sigma} \sum_{n=1}^{N} n^{-1} \approx \frac{1}{\sigma}
\left[\, \ln N + \gamma \, \right]
\label{wsq}
\end{equation}
where $\gamma=0.577...$ is the Euler constant, while
the fluctuations of $w_2$ are finite
\begin{equation}
\langle w_2^2 \rangle -\langle w_2 \rangle^2
= \frac{a^2}{\sigma^2} \,\, ; \hspace{0.5truecm} a=\frac{\pi}{\sqrt{6}} \, .
\label{wssq}
\end{equation}
This means that if we scale $w_2$ with $\langle w_2 \rangle$ then the width
of the distribution goes to zero for $N\to \infty$ hence the conclusion
about the delta function. In order not to lose information about the possible
structure at small $w_2-\langle w_2 \rangle$, one should introduce a
scaling variable which expands the delta function. This can be achieved
by introducing the following variable \cite{HoldsXY1}
\begin{equation}
y= \frac{w_2-\langle w_2 \rangle}{\sqrt{\langle w_2^2 \rangle-\langle w_2
\rangle^2}} \, .
\label{scalvar}
\end{equation}
Substituting $w_2$ from the above expression into Eq.(\ref{Pwsc1}) and using
Eqs.(\ref{wsq}) and (\ref{wssq}), one finds that
the limit $N\to \infty$ can be taken and a scaling function in terms of
$y$ emerges
\begin{equation}
\Phi(y)\equiv\sqrt{\langle w_2^2 \rangle -\langle w_2 \rangle^2}P(w_2) =
\int\limits_{-i\infty}^{i\infty} \frac{ds}{2 \pi i}
\,e^{x s} \prod_{n=1}^{\infty}
\frac{ e^{\frac{s}{a n}}}{1+{\frac{s}{a n}}}
\label{Pwsc} \, .
\end{equation}
Noting that infinite product above can be expressed \cite{Abra-Steg} through the
$\Gamma$ function
\begin{equation}
\prod_{n=1}^{\infty}
\frac{ e^{\frac{s}{a n}}}{1+{\frac{s}{a n}}}=e^{\gamma s/a}\Gamma\left(1+\frac{s}{a}\right)
\label{Pwsc3} \, ,
\end{equation}
and using
Euler's integral representation for the $\Gamma$ function, one finds a surprisingly simple
analytical result
\begin{equation}
  \Phi(y)=ae^{-(ay+\gamma) -e^{-(ay+\gamma)}} \, .
\label{Gumbeldist}
\end{equation}
The real surprise actually is that $\Phi(y)$, shown in Fig.~\ref{fig:Gumbel}, is the
the so called Fisher-Tippett-Gumbel
distribution \cite{Gumbel} which is one of the three limiting forms of
extreme value statistics.
   \begin{figure}[ht]
   \begin{center}
   \vspace{0.3cm}
   \begin{tabular}{c}
   \includegraphics[height=5.5cm]{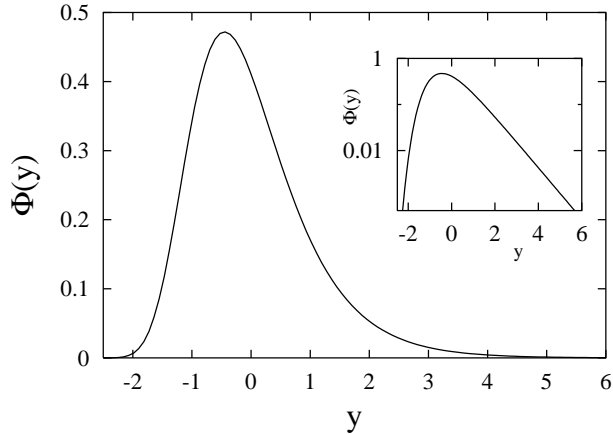}
   \end{tabular}
   \end{center}
   \vspace{-0.3cm}
   \caption[example]
   { \label{fig:Gumbel}
   The Fisher-Tippett-Gumbel distribution. The inset shows it on semilogarithmic
   scale to demonstrate the exponential decay at large arguments.}
   \end{figure}

It should be emphasized that we do not see any physical reason why should there
be a connection between the fluctuations of a $1/f$ signal and the extreme value
distributions. It seems to be a puzzle, however, whose solution may reveal an interesting
picture.

\section{Effect of boundary conditions}
\label{sect:pi}

As a final topic, I would like to consider the boundary-condition dependence of
the scaling functions. As discussed in previous
sections, the universality is due to the diverging fluctuations which in turn imply that
the correlation length is infinite in the system. The infinite correlations
mean that the boundaries are felt in the bulk
and thus it is not entirely unexpected that the scaling functions
are sensitive to the boundaries. The important question is how large the
boundary effects are.

The effects of boundaries have been studied \cite{1falpha} for
Gaussian signals with $1/f^\alpha$ power spectrum. It was found that the
boundary effects are large for $\alpha>4$, easily noticeable in the
range $1<\alpha<4$, and they disappear entirely for $\alpha<0.5$.
In order to demonstrate the magnitude of the boundary effect
in the physically most relevant range of
$1<\alpha<4$, I would like to consider the problem of the randomness of the
digits of $\pi$ which, as we shall see,
corresponds to an $\alpha=2$ problem.

The statistical properties of the digits of $\pi$ appears to
occupy the mind of a few
mathematicians{ \cite{Borwein-pi}~\footnote[1]{See e.g. http://www.sfu.ca/\~~pborwein}}.
Although no rigorous proof exists yet, it is believed that the digits
are random. This belief is based on generating and analyzing a large number $(10^7-10^8)$
of digits. We shall test the belief of randomness by mapping the digits onto an interface
(see Fig.\ref{fig:pi-rooftop}) and calculating the scaling function of the
width of the interface.

   \begin{figure}[ht]
   \begin{center}
   \begin{tabular}{c}
   \includegraphics[height=5.5cm]{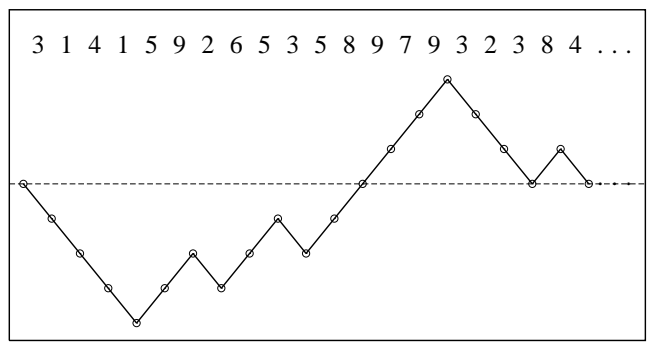}
   \end{tabular}
   \end{center}
   \vspace{-0.8cm}
   \caption[example]
   { \label{fig:pi-rooftop}
   Mapping between the digits of $\pi$ and a configuration of the so called rooftop
   model \cite{Meakin-rooftop,PRL-rooftop} of interface.
   The slope of the interface can be only $\pm 1$ with $-1$
   occurring for digits $0-4$ and $+1$ for $5-9$.}
   \end{figure}

If the digits of $\pi$ are random then the interface shown in Fig.\ref{fig:pi-rooftop}
is just a random walk
equivalent to the steady state of the Edwards-Wilkinson model as well as to
Gaussian signals with $1/f^2$ power spectrum. For periodic $1/f^2$ signals the
width distribution was calculated in Ref.~\cite{FORWZ}. It is not straightforward,
however, to make a comparison with this distribution function
since the digits of $\pi$ provide only a long but single
signal (not unlike to some experimental situations). A histogram of roughness can be
built, nevertheless, by calculating $w_2$ for segments (windows)
of the signal whose size is much smaller than the total
length but at the same time large enough that the finite-size effects
in the scaling function would be small. Clearly, the boundary conditions for the segments
are not periodic (we call them  window boundary conditions,
WBC). The PDF of the roughness using WBC turns out to be a well defined function
in the limit of the total length of the signal going to infinity,
and, furthermore, this function is found \cite{1falpha} to be independent of the
boundary conditions for the full signal. Comparison of this function with the
width distribution of $\pi$ is displayed on Fig.\ref{fig:pi-BC2} where one can also
see the scaling function for periodic boundary conditions (PBC). The difference between
the PBC and WBC scaling functions is easily observable and it is clear that we would have
concluded incorrectly about the non randomness of the digits of $\pi$
had we ignored the boundary condition dependence of the scaling functions.

   \begin{figure}[ht]
   \vspace{1truecm}
   \begin{center}
   \begin{tabular}{c}
   \includegraphics[height=6cm]{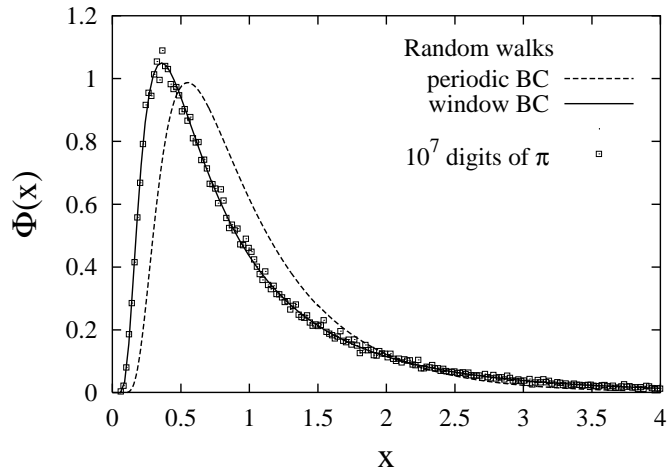}
   \end{tabular}
   \end{center}
   \caption[example]
   { \label{fig:pi-BC2}
   Distribution of the roughness of $\pi$ calculated by mapping its
   first $10^7$ digits onto the rooftop interface model (Fig.\ref{fig:pi-rooftop}).
   The scaled distribution $\Phi=<w_2>P(w_2)$ is obtained by
   using a measuring window of $10^3$ digits and the
   scaled variable is $x=w_2/<w_2>$.
   The comparison is with
   random walks with periodic- and window (free) boundary conditions. The collapse
   with the window BC curve is in agreement with the notion that the
   digits of $\pi$ are random.}
   \end{figure}

The moral from the story of $\pi$ is twofold. On one hand it cautions us that
the scaling functions are useful theoretical instruments only if
the boundary conditions for experimental
data are carefully specified both at the measuring and at the analyzing stage.
On the other hand, the boundary-condition dependence of the scaling functions
suggests that they "{\it hear the shape of the drum}" i.e. they may be used
to see the shape of objects which are embedded in other media and
can be seen only through their fluctuations.

\section{Final remarks}
It should be clear that only a small part of the picture gallery was
exhibited in this lecture. The gallery is far from
complete and we expect that scaling functions originating from newly discovered
nonequilibrium universality classes will enrich it regularly.
Furthermore, we expect that limiting
distributions such as the ones emerging in extreme statistics will also
be included and will have much wider use in physics.
Finally, strongly fluctuating quantum systems may also provide valuable novelties.
\acknowledgments

I am grateful to my colleagues who shaped my ideas on
nonequilibrium distribution functions and
collaborated \cite{FORWZ,PRZ94,RP94,AR96,1f,1falpha,KPZ-width,Eisler2003}
in the creation of the picture gallery.
This work has been partially supported by the Hungarian Academy
of Sciences (Grants No. OTKA T029792 and T043734).

\end{document}